\documentclass
[floats,floatfix,showpacs,preprint,prd,superscriptaddress,onecolumn,nofootinbib]{revtex4}%
\usepackage{amsmath, amsfonts}
\usepackage{bm}
\usepackage{color}
\usepackage{amsmath}
\usepackage{amsfonts}
\usepackage{amssymb}
\usepackage{graphicx}%
\providecommand{\U}[1]{\protect\rule{.1in}{.1in}}

\newcommand{\be}{\begin{equation}}
\newcommand{\ee}{\end{equation}}
\newcommand{\bea}{\begin{eqnarray}}
\newcommand{\eea}{\end{eqnarray}}

\begin{document}
\title{Quintic quasi-topological gravity}
\author{Adolfo Cisterna}
\email{adolfo.cisterna@ucentral.cl}
\affiliation{Universidad Central de Chile, Vicerrector\'{\i}a acad\'{e}mica, Toesca 1783
Santiago, Chile.}
\affiliation{Instituto de Ciencias F\'{\i}sicas y Matem\'{a}ticas, Universidad Austral de
Chile, Casilla 567, Valdivia, Chile}
\author{Luis Guajardo}
\email{luis.guajardo.r@gmail.com}
\affiliation{Instituto de Matem\'{a}tica y F\'{\i}sica, Universidad de Talca, Casilla 747,
Talca, Chile}
\author{Mokhtar Hassa\"{\i}ne}
\email{hassaine@inst-mat.utalca.cl}
\affiliation{Instituto de Matem\'{a}tica y F\'{\i}sica, Universidad de Talca, Casilla 747,
Talca, Chile}
\author{Julio Oliva}
\email{julioolivazapata@gmail.com}
\affiliation{Departamento de F\'{\i}sica, Universidad de Concepci\'{o}n, Casilla, 160-C,
Concepci\'{o}n, Chile}

\begin{abstract}
We construct a quintic quasi-topological gravity in five dimensions, i.e. a
theory with a Lagrangian containing $\mathcal{R}^5$ terms and whose field
equations are of second order on spherically (hyperbolic or planar) symmetric
spacetimes. These theories have recently received attention since when
formulated on asymptotically AdS spacetimes might provide for gravity duals of
a broad class of CFTs. For simplicity we focus on five dimensions. We show
that this theory fulfils a Birkhoff's Theorem as it is the case in Lovelock
gravity and therefore, for generic values of the couplings, there is no
$s$-wave propagating mode. We prove that the spherically symmetric solution is
determined by a quintic algebraic polynomial equation which resembles
Wheeler's polynomial of Lovelock gravity. For the black hole solutions we
compute the temperature, mass and entropy and show that the first law of black
holes thermodynamics is fulfilled. Besides of being of fourth order in
general, we show that the field equations, when linearized around AdS are of
second order, and therefore the theory does not propagate ghosts around this background.
Besides the class of theories originally introduced in arXiv:1003.4773
[gr-qc], the general geometric structure of these Lagrangians remains an open problem.

\end{abstract}
\maketitle

\section{Introduction}

The geometric classical description of the gravitational interaction naturally
leads to General Relativity as the unique, diffeomorphism invariant theory
with second order field equations in four dimensions. In spite of the
successes the theory has had, the lack of a proper quantum description of
gravitational phenomena led to the exploration of alternative scenarios which
have received vast attention during the last decades. If one requires some
level of predictability, such alternative scenarios might be restricted in
some manner. As before, requirements such as diffeomorphism invariance and
second order field equations are usually considered cornerstones in this
process that naturally leads to Einstein-Hilbert action in three and four
dimensions and to Lovelock theories in arbitrary dimension $D$
\cite{Lovelock:1971yv}. When formulated on asymptotically AdS spacetimes,
using the tools of AdS/CFT correspondence \cite{AdSCFT} one can treat these
theories as gravity duals of some Conformal Field Theory living at the
boundary. As in the low energy limit of string theory, besides the
Einstein-Hilbert action for gravity, higher curvature corrections might appear
in a perturbative treatment, which can differ in general from those in the
Lovelock family. Nevertheless one can hope that some of the physics of these
higher curvature corrections could be correctly captured by their Lovelock
counterpart. These kind of explorations led for example to understand that
finite contributions from quadratic terms might induce violations of the
$\eta/s$ KSS bound \cite{KSS} (see \cite{VKSS1} and \cite{VKSS2}). Lovelock
theories provide one with a setup where certain control at a computational
level can be attained. More precisely, in Lovelock theories one can find exact
analytic static black holes which provide for finite temperature duals with
interesting thermal properties and phase diagram structure. Within Lovelock
family in five dimensions, General Relativity is supplemented by a single
quadratic term (which is the dimensional continuation of the four dimensional
Euler density for Euclidean compact manifolds without boundary) and besides
Newton's constant, the theory has an extra dimensionful coupling. In the
context of the AdS$_{5}$/CFT$_{4}$ correspondence departing from this family
would allow the dual CFT to be non-supersymmetric \cite{Hofman:2008ar}, at the
cost of loosing some of the analytic computational control. A natural question
therefore arises: Is it possible to define a sensible gravity theory in five
dimensions beyond the Einstein-Gauss-Bonnet theory? The so-called
quasi-topological gravities provide for such an example. In reference
\cite{OR} a new cubic gravity theory in five dimensions was introduced. The
cubic combination reads
\begin{multline}
\mathcal{L}_{3}=-\frac{7}{6}R_{\ \ cd}^{ab}R_{\ \ bf}^{ce}R_{\ \ ae}%
^{df}-R_{ab}^{\ \ cd}R_{cd}^{\ \ be}R_{\ e}^{a}-\frac{1}{2}R_{ab}%
^{\ \ cd}R_{\ c}^{a}R_{\ d}^{b}\\
+\frac{1}{3}R_{\ b}^{a}R_{\ c}^{b}R_{\ a}^{c}-\frac{1}{2}RR_{\ b}^{a}%
R_{\ a}^{b}+\frac{1}{12}R^{3}\ , \label{L3OR}%
\end{multline}
and can be singled out as the unique cubic combination in five dimensions
whose traced field equations lead to a second order constraint and that also
has second order field equation on general spherically (planar or hyperbolic)
symmetric spacetimes \cite{ORclassif}\footnote{The quasi-topological
combination found in \cite{MyersRobinson} $\mathcal{Z}^{\prime}$, is related
to $\mathcal{L}_{3}$ by the relation $\mathcal{L}_{3}=\frac{7}{24}%
\mathcal{Z}^{\prime}+\frac{7}{48}\mathcal{E}_{6}$, where $\mathcal{E}_{6}$
stands for the six-dimensional Euler that vanishes identically in five
dimensions.}. It was also realized in \cite{OR} that this theory belongs to a
general family of Lagrangians of order $k$ in the curvature, that can be
constructed in dimensions $D=2k-1$ for $k\geq3$, and have the simple form%
\begin{multline}
\mathcal{\tilde{L}}_{k}={\frac{1}{2^{k}}}\left(  \frac{1}{D-2k+1}\right)
\delta_{c_{1}d_{1}\cdots c_{k}d_{k}}^{a_{1}b_{1}\cdots a_{k}b_{k}}\left(
C_{a_{1}b_{1}}^{c_{1}d_{1}}\cdots C_{a_{k}b_{k}}^{c_{k}d_{k}}-R_{a_{1}b_{1}%
}^{c_{1}d_{1}}\cdots R_{a_{k}b_{k}}^{c_{k}d_{k}}\right) \\
-c_{k}C_{a_{1}b_{1}}^{a_{k}b_{k}}C_{a_{2}b_{2}}^{a_{1}b_{1}}\cdots
C_{a_{k}b_{k}}^{a_{k-1}b_{k-1}}\ . \label{Lk}%
\end{multline}
Here $C_{a\ cd}^{\ b}$ is the Weyl tensor and
\begin{equation}
c_{k}={\frac{(D-4)!}{(D-2k+1)!}}{\frac
{[k(k-2)D(D-3)+k(k+1)(D-3)+(D-2k)(D-2k-1)]}{[(D-3)^{k-1}(D-2)^{k-1}%
+2^{k-1}-2(3-D)^{k-1}]}\ .}%
\end{equation}
The Lagrangian (\ref{L3OR}) is obtained by setting $D=5$ and $k=3$ in
(\ref{Lk}) after expanding the Weyl tensor in terms of Riemann tensor and its
traces. It's also interesting to note that after expanding the Lagrangian
(\ref{Lk}) in the case $k=2$, and then setting $D=3$, leads to the quadratic
part of the New Massive Gravity Lagrangian \cite{Bergshoeff:2009hq}.

The $\mathcal{\tilde{L}}_{k\geq3}$ expression in (\ref{Lk}) allows to directly
verify that the linearized field equations around any conformally flat
background ((A)dS in particular) are of second order. On the other hand, the
relative factor $c_{k}$ in (\ref{Lk}) is such that on spacetimes that are
conformal to spherically (hyperbolic or planar) symmetric spacetimes, the
higher derivative contributions to the field equations coming from the
variations of the Weyl tensors, cancel each other. This is possible since for
such family of spacetimes, all the component of the Weyl tensor $C_{\ \ cd}%
^{ab}$ are proportional to a single function $X$ \cite{DeserRyzhov}, and
therefore any scalar constructed as a complete contraction of $k$ Weyl tensors
will be proportional to $X^{k}$ (see \cite{ORBirkhoff1} and \cite{ORBirkhoff2}
for related further developments). These results allow to construct a new
theory, quartic in the Riemann, in $D=7$.

Motivated by broadening the family of four-dimensional CFTs with relatively
simple gravity duals, in reference \cite{originalquartic} the authors
successfully look for a quartic theory with second order field equations on
spherical symmetry. Such theory was dubbed Quartic quasi-topological gravity
and when supplemented with terms of lower order in the Riemann tensor, leads
to second order linearized field equations around AdS spacetime. A property
that is shared by its cubic counterpart. It's important to notice that quartic
quasi-topological gravity in five dimensions does not belong to the family
defined in equation (\ref{Lk}), nevertheless on spherically symmetric
spacetimes, the field equations are simple and reduce to a generalized
Wheeler-like polynomial equation for the lapse function \cite{Wheeler:1985qd}
(see also \cite{bestiary}). The authors of \cite{originalquartic} also
conjecture the existence of quasi-topological gravities of arbitrarily high
degree in the curvature. The purpose of the present paper is to show that for
Lagrangians that are quintic in the curvature, $\mathcal{R}^{5}$, this is
indeed the case.

\bigskip

The paper is organized as follows: In Section II we show that there is at
least one particular combination of quintic algebraic invariants in five
dimension which leads to second order field equations on spherically symmetric
spacetimes. In Section III we present the quintic algebraic Wheeler's
polynomial that determines the black hole solutions of the theory, we
characterize the maximally symmetric solutions and show that for generic
values of the couplings, the asymptotic behavior of the solutions coincides
with the one of GR. Section IV is devoted to the proof of Birkhoff's theorem
in this setup which shows that for generic values of the couplings of a
non-homogeneous curvature combination, the theory does not propagate $s$-waves
in spite of being in general a higher derivative theory. Due to the
Abel-Rufini theorem the obtained Wheeler's-like polynomial cannot be solved by
radicals, nevertheless, assuming the existence of an event horizon, we are
able to compute the temperature, entropy and mass of the black hole solutions
in Section V. Section VI contains the proof that the theory is ghosts-free
around AdS. In section VII we analyze some particular cases of the values of
the couplings that allow for an explicit solution of the quintic polynomial
and at the same time lead to interesting black hole solutions. Section VIII
contains further remarks and conclusions.

\section{The theory}

Here we consider the following gravity theory%
\begin{equation}
I\left[  g_{\mu\nu}\right]  =\int\sqrt{-g}d^{5}x\left[  \frac{R-2\Lambda
}{16\pi G}+\sum_{k=2}^{5}a_{k}\mathcal{L}_{k}\right]  \ ,\label{fullaction}%
\end{equation}
where $\mathcal{L}_{2}$ stands for the Gauss-Bonnet combination%
\begin{equation}
\mathcal{L}_{2}:=R^{2}-4R_{ab}R^{ab}+R_{abcd}R^{abcd}\ ,
\end{equation}
the cubic $\mathcal{L}_{3}$ term is the cubic quasi-topological combination in
(\ref{L3OR}) and $\mathcal{L}_{4}$ stands for the quartic quasi-topological
term that can be written as
\begin{align}
\mathcal{L}_{4} &  =\frac{1}{73\times2^{5}\times3^{2}}\left[  7080R^{pqbs}%
R_{p\ b}^{\ a\ u}R_{a\ u}^{\ v\ w}R_{qvsw}-234R^{pqbs}R_{pq}^{\ \ au}%
R_{au}^{\ \ vw}R_{bsvw}-1237\left(  R^{pqbs}R_{pqbs}\right)  ^{2}\right.
\nonumber\\
&  +1216R^{pq}R^{bsau}R_{bs\ p}^{\ \ v}R_{auvq}-6912R^{pq}R^{bs}%
R_{\ p\ q}^{a\ u}R_{abus}-7152R^{pq}R^{bs}R_{\ \ pb}^{au}R_{auqs}\nonumber\\
&  +308R^{pq}R_{pq}R^{bsau}R_{bsau}+298R^{2}R^{pqbs}R_{pqbs}+12864R^{pq}%
R^{bs}R_{b}^{\ a}R_{psqa}-115R^{4}\nonumber\\
&  \left.  -912RR^{pq}R^{bs}R_{pbqs}+4112R^{pq}R_{p}^{\ b}R_{q}^{\ s}%
R_{bs}-4256RR^{pq}R_{p}^{\ b}R_{qb}+1156R^{2}R^{pq}R_{pq}\right]
\ .\label{PoneteUnaCorbata2}%
\end{align}

The new quintic quasi-topological combination is%
\begin{align}
&  \mathcal{L}_{5}=A_{1}RR_{b}^{\ a}R_{c}^{\ b}R_{d}^{\ c}R_{a}^{\ d}%
+A_{2}RR_{b}^{\ a}R_{a}^{\ b}R_{ef}^{\ \ cd}R_{cd}^{\ \ ef}+A_{3}RR_{c}%
^{\ a}R_{d}^{\ b}R_{ef}^{\ \ cd}R_{ab}^{\ \ ef}\nonumber\\
&  +A_{4}R_{b}^{\ a}R_{a}^{\ b}R_{d}^{\ c}R_{e}^{\ d}R_{c}^{\ e}+A_{5}%
R_{b}^{\ a}R_{c}^{\ b}R_{a}^{\ c}R_{fg}^{\ \ de}R_{de}^{\ \ fg}+A_{6}%
R_{b}^{\ a}R_{d}^{\ b}R_{f}^{\ c}R_{ag}^{\ \ de}R_{ce}^{\ \ fg}\nonumber\\
&  +A_{7}R_{b}^{\ a}R_{d}^{\ b}R_{f}^{\ c}R_{cg}^{\ \ de}R_{ae}^{\ \ fg}%
+A_{8}R_{b}^{\ a}R_{c}^{\ b}R_{ae}^{\ \ cd}R_{gh}^{\ \ ef}R_{df}%
^{\ \ gh}+A_{9}R_{b}^{\ a}R_{c}^{\ b}R_{ef}^{\ \ cd}R_{gh}^{\ \ ef}%
R_{ad}^{\ \ gh}\nonumber\\
&  +A_{10}R_{b}^{\ a}R_{c}^{\ b}R_{eg}^{\ \ cd}R_{ah}^{\ \ ef}R_{df}%
^{\ \ gh}+A_{11}R_{c}^{\ a}R_{d}^{\ b}R_{ab}^{\ \ cd}R_{gh}^{\ \ ef}%
R_{ef}^{\ \ gh}+A_{12}R_{c}^{\ a}R_{d}^{\ b}R_{ae}^{\ \ cd}R_{gh}%
^{\ \ ef}R_{bf}^{\ \ gh}\nonumber\\
&  +A_{13}R_{c}^{\ a}R_{d}^{\ b}R_{ef}^{\ \ cd}R_{gh}^{\ \ ef}R_{ab}%
^{\ \ gh}+A_{14}R_{c}^{\ a}R_{d}^{\ b}R_{eg}^{\ \ cd}R_{ah}^{\ \ ef}%
R_{bf}^{\ \ gh}+A_{15}R_{c}^{\ a}R_{e}^{\ b}R_{af}^{\ \ cd}R_{gh}%
^{\ \ ef}R_{bd}^{\ \ gh}\nonumber\\
&  +A_{16}R_{b}^{\ a}R_{ad}^{\ \ bc}R_{fh}^{\ \ de}R_{ci}^{\ \ fg}%
R_{eg}^{\ \ hi}+A_{17}R_{b}^{\ a}R_{de}^{\ \ bc}R_{cf}^{\ \ de}R_{hi}%
^{\ \ fg}R_{ag}^{\ \ hi}+A_{18}R_{b}^{\ a}R_{df}^{\ \ bc}R_{ac}^{\ \ de}%
R_{hi}^{\ \ fg}R_{eg}^{\ \ hi}\nonumber\\
&  +A_{19}R_{b}^{\ a}R_{df}^{\ \ bc}R_{ah}^{\ \ de}R_{ei}^{\ \ fg}%
R_{cg}^{\ \ hi}+A_{20}R_{b}^{\ a}R_{df}^{\ \ bc}R_{gh}^{\ \ de}R_{ei}%
^{\ \ fg}R_{ac}^{\ \ hi}+A_{21}R_{cd}^{\ \ ab}R_{eg}^{\ \ cd}R_{ai}%
^{\ \ ef}R_{fj}^{\ \ gh}R_{bh}^{\ \ ij}\nonumber\\
&  +A_{22}R_{ce}^{\ \ ab}R_{af}^{\ \ cd}R_{gi}^{\ \ ef}R_{bj}^{\ \ gh}%
R_{dh}^{\ \ ij}+A_{23}R_{ce}^{\ \ ab}R_{ag}^{\ \ cd}R_{bi}^{\ \ ef}%
R_{fj}^{\ \ gh}R_{dh}^{\ \ ij}+A_{24}R_{ce}^{\ \ ab}R_{fg}^{\ \ cd}%
R_{hi}^{\ \ ef}R_{aj}^{\ \ gh}R_{bd}^{\ \ ij}\ , \label{uff}%
\end{align}
where the (not illuminating at all) $A_{k}$ coefficients, are defined in the
appendix. The dimensionful couplings $a_{k}$ have mass dimension $5-2k$.

The quasi-topological combinations seem to be very cumbersome but the field
equations, on spherically symmetric spacetimes reduce to very simple expressions.

Clearly, the quasi-topological gravities are defined up to the addition of the
corresponding Euler densities that in the cubic, quartic and quintic cases,
identically vanish in five dimensions. A different set of identities that can
be used to provide a different expression for these Lagrangians make use of
the cubic and quartic Lovelock tensor $\mathcal{G}_{\mu\nu}^{\left(  3\right)
}$ and $\mathcal{G}_{\mu\nu}^{\left(  4\right)  }$. These tensors identically
vanish in dimension five and can be used to construct new scalar identities by
contracting them with symmetric curvature combinations, as well as with
curvature scalars. Beyond these trivial sources of non-uniqueness for the
explicit form of the quasi-topological Lagrangians, there is a more subtle one
that leads to theories that are intrinsically different. Since the quartic and
quintic theories are defined by requiring second order field equations on
spherically symmetric spacetimes, a combination that identically vanishes for
this class of spacetimes can be added. The differences between the theories so
defined will become manifest only when they are explored beyond spherical
symmetry. An example of this is provided by the addition of a general
combination of the independent complete contractions of $k$ Weyl tensor
$Tr_{\left(  p\right)  }\left(  C^{k}\right)  $, where $Tr_{\left(  p\right)
}$ stands for a particular way of pairing the indices of the different Weyl
tensors. When evaluated on spacetimes that are conformal to spherically
symmetric ones, the different traces $Tr_{\left(  p\right)  }$ turn out to be
proportional \cite{DeserRyzhov}. Therefore a combination of these traces,
fixed by a single constraint, can be added to the quasi-topological terms to
provide new theories which on spherical symmetry coincide (see the discussion
above equation (88) in \cite{OR}). As mentioned in \cite{ORclassif}, since
there are only two possible cubic complete contractions of Weyl tensors,
$Tr_{\left(  1\right)  }\left(  C^{3}\right)  $ and $Tr_{\left(  2\right)
}\left(  C^{3}\right)  $, cubic quasi-topological gravity is unique. Quartic
quasi-topological gravity is also argued to be unique in the original
reference \cite{originalquartic}. Whether or not the quintic quasi-topological
theory here presented is unique, goes beyond the scope of this work and it
would require at least to properly classify all the non-trivial, independent
traces of the form $Tr_{\left(  p\right)  }\left(  C^{5}\right)  $. In ten
dimensions, these are the terms that define the $c$-contributions of the
conformal anomaly (see e.g. \cite{Deser:1993yx} and \cite{Boulanger:2007ab}).

\bigskip

Hereafter we set $16\pi G=1$.

\section{Static solutions}

Wheeler's polynomial for a static metric of the form%
\begin{equation}
ds^{2}=-f\left(  r\right)  dt^{2}+\frac{dr^{2}}{g\left(  r\right)  }%
+r^{2}d\Sigma_{\gamma}^{2}\ , \label{static}%
\end{equation}
with $f\left(  r\right)  =g\left(  r\right)  $, reads%
\begin{equation}
\left(  f-\gamma\right)  ^{5}\frac{a_{5}}{r^{6}}-\left(  f-\gamma\right)
^{4}\frac{a_{4}}{r^{4}}-2\left(  f-\gamma\right)  ^{3}\frac{a_{3}}{r^{2}%
}-12\left(  f-\gamma\right)  ^{2}a_{2}+6r^{2}\left(  f-\gamma\right)
+r^{4}\Lambda=-\mu\ . \label{wheeler}%
\end{equation}
Where $d\Sigma_{\gamma}$ denotes the line element of a Euclidean three
dimensional manifold of normalized constant curvature $\gamma=\pm1$ or $0$,
and $\mu$ is an integration constant, that will determine the mass of the
solution. This is a quintic algebraic equation that due to Abel-Rufini
theorem, cannot be solved in general by radicals. We will show in the next
section that for generic values of the couplings the condition $f\left(
r\right)  =g\left(  r\right)  $ is actually an output of the field equations
and not a restriction put by hand.

Let's assume that the maximally symmetric solution has a dressed constant
curvature%
\begin{equation}
R_{\ \ \beta\sigma}^{\mu\nu}=\frac{\lambda}{6}\left(  \delta_{\beta}^{\mu
}\delta_{\sigma}^{\nu}-\delta_{\sigma}^{\nu}\delta_{\beta}^{\mu}\right)  \ ,
\end{equation}
whose metric can be represented by setting%
\begin{equation}
f\left(  r\right)  =g\left(  r\right)  =-\frac{\lambda}{6}r^{2}+\gamma\ .
\end{equation}
on the metric (\ref{static}). The dressed curvature is fixed by the following
polynomial%
\begin{equation}
P\left[  \lambda\right]  :=a_{5}\lambda^{5}+6a_{4}\lambda^{4}-72a_{3}%
\lambda^{3}+2592a_{2}\lambda^{2}+7776\left(  \lambda-\Lambda\right)  =0\ .
\label{PTLambda}%
\end{equation}
\newline As usual, depending on the values of the couplings $\left(
a_{5},a_{4},a_{3},a_{2},\Lambda\right)  $ one can have from one to five
different vacua, and one can identify the G.R. branch, as the one for which
$\lambda\rightarrow\Lambda$ as $a_{2}$, $a_{3}$, $a_{4}$ and $a_{5}$ go to zero.

\bigskip

It's possible to show that for generic values of the couplings, the asymptotic
behavior allowed by Wheeler's polynomial (\ref{wheeler}), coincides with that
of General Relativity. In fact assuming%
\begin{equation}
f\left(  r\right)  =g\left(  r\right)  =-\frac{\lambda}{6}r^{2}+\gamma
+j\left(  r\right)  \ ,
\end{equation}
with $j\left(  r\right)  $ having a Laurent expansion at infinity that does
not modify the leading term of the metric, i.e.%
\begin{equation}
j\left(  r\right)  =c_{1}r+c_{0}+\frac{c_{2}}{r}+\frac{c_{3}}{r^{2}%
}+\mathcal{O}\left(  r^{-3}\right)  \ ,
\end{equation}
one can expand (\ref{wheeler}) at infinity and solve it order by order. This
leads to the following set of conditions%
\begin{equation}
P\left[  \lambda\right]  r^{4}=0\ ,\ c_{1}\frac{dP\left[  \lambda\right]
}{d\lambda}r^{3}=0\ \ , \label{r2rasymp}%
\end{equation}
where $P\left[  \lambda\right]  $ is defined in (\ref{PTLambda}). Once
$P\left[  \lambda\right]  $ is solved, which fixes the value of the dressed
cosmological constant $\lambda$, requiring at the same time $\frac{dP\left[
\lambda\right]  }{d\lambda}$ to vanish, would imply a relation between the
couplings, therefore for generic values of the latter, (\ref{r2rasymp}) imply
that $c_{1}$ has to vanish. Considering this on the next relevant order of the
asymptotic expansion of Wheeler's polynomial one obtains%
\begin{equation}
c_{0}\frac{dP\left[  \lambda\right]  }{d\lambda}r^{2}=0\ ,
\end{equation}
which again implies that generically $c_{0}=0$. Then one obtains $c_{2}%
\frac{dP\left[  \lambda\right]  }{d\lambda}r=0$, which forces to set $c_{2}%
=0$, and finally one obtains%
\begin{equation}
c_{3}\frac{dP\left[  \lambda\right]  }{d\lambda}+1296\mu=0\ .
\end{equation}
As mentioned before, this implies that the asymptotic behavior of the full
solution matches the one of General Relativity since%
\begin{equation}
f\left(  r\right)  =g\left(  r\right)  =-\frac{\lambda}{6}r^{2}+\gamma
-\frac{\mu}{1296\frac{dP\left[  \lambda\right]  }{d\lambda}}\frac{1}{r^{2}%
}+\mathcal{O}\left(  r^{-3}\right)  \ .
\end{equation}
This shows that on spherically symmetric spacetimes, the only way of relaxing
the asymptotic conditions of quasi-topological gravity (with respect to those
of G.R.), would be to consider the critical cases in which the polynomial
$P\left[  \lambda\right]  $ has roots with multiplicity larger than one. In
Section VI we consider some particular regions of the space of couplings that
allow for relaxed asymptotic behavior.

\section{Birkhoff's theorem}

It's well known that the field equations of a gravitational theory on
spherically symmetric spacetimes, are correctly reproduced from the variation
of a reduced action \cite{Palais:1979rca},\cite{Deser:2004gi}, obtained from
the evaluation of the Lagrangian on the metric%
\begin{equation}
ds^{2}=-a\left(  t,r\right)  b^{2}\left(  t,r\right)  dt^{2}+2f\left(
t,r\right)  b\left(  t,r\right)  dtdr+\frac{dr^{2}}{a\left(  t,r\right)
}+r^{2}d\Sigma_{\gamma}^{2}\ , \label{minisuptimedep}%
\end{equation}
where $d\Sigma_{\gamma}$ denotes the line element of a Euclidean three
dimensional manifold of constant curvature $\gamma=\pm1$ or $0$. Evaluating
the action (\ref{fullaction}) on (\ref{minisuptimedep}) one obtains a reduced
action which is a functional of $I\left[  a,b,f\right]  $. It's convenient to
introduce $h\left(  t,r\right)  $ such that%
\begin{equation}
a\left(  t,r\right)  =\gamma+h\left(  t,r\right)  \ .
\end{equation}
The variation of the reduced action with respect to $a$, $b$ and $f$, and a
posteriori gauge fixing $f=0$, respectively lead to the field equations%
\begin{align}
0  &  =\left(  -24r^{6}h(t,r)a_{2}-6h(t,r)^{2}r^{4}a_{3}-4h(t,r)^{3}r^{2}%
a_{4}+5h(t,r)^{4}a_{5}+6r^{8}\right)  \frac{\partial b\left(  t,r\right)
}{\partial r}\label{birka}\\
0  &  =h(t,r)^{5}r^{-5}a_{5}-h(t,r)^{4}r^{-3}a_{4}-2h(t,r)^{3}r^{-1}%
a_{3}-12h(t,r)^{2}ra_{2}\nonumber\\
&
\ \ \ \ \ \ \ \ \ \ \ \ \ \ \ \ \ \ \ \ \ \ \ \ \ \ \ \ \ \ \ \ \ \ \ \ \ \ \ \ \ \ \ \ \ \ \ \ \ \ \ \ \ \ \ \ \ \ \ \ \ +r^{5}%
\Lambda+6r^{3}h(t,r)+\mu(t)r\label{birkb}\\
0  &  =\left(  -24r^{6}h(t,r)a_{2}-6h(t,r)^{2}r^{4}a_{3}-4h(t,r)^{3}r^{2}%
a_{4}+5h(t,r)^{4}a_{5}+6r^{8}\right)  \frac{\partial h\left(  t,r\right)
}{\partial t} \label{birkf}%
\end{align}
The function $f\left(  t,r\right)  $ appears as a Lagrange multiplier on the
reduced action and provides one with the $(t,r)$ component of the field
equations. Here $\mu\left(  t\right)  $ is an integration function. For
generic values of the couplings, provided (\ref{birkb}) is fulfilled, one can
show that the pre-factors of the derivatives in equations (\ref{birka}) and
(\ref{birkf}) are non-vanishing and therefore $b\left(  t,r\right)  =b\left(
t\right)  $ and $\frac{\partial h}{\partial t}=0$. The latter implies
$\mu\left(  t\right)  =\mu$. As usual the function $b\left(  t\right)  $ can
be absorbed by a time reparametrization. This proves the Birkhoff's theorem
for the non-homogenous combination, i.e., for generic values of the couplings,
the spherically (planar or hyperbolic) symmetric solution is static and is
determined by a quintic polynomial equation. These results are analogous to
the Birkhoff's theorem in Lovelock gravity (see \cite{Zegers:2005vx} and
\cite{Deser:2005gr}). It would be interesting to explore whether the
generalized Birkhoff's theorem of Chern-Simons \cite{Oliva:2012ff} and
Lovelock theory \cite{Dotti:2010bw}-\cite{Ray:2015ava} can be extended for the
quasi-topological theories, for general horizon geometries.

\section{Black hole thermodynamics}

Assuming that there is a non-degenerate horizon, that is, assuming that the
polynomial equation (\ref{wheeler}) has a solution $f(r)$ with a single zero
located at $r=r_{h}$, we now analyze the thermodynamical properties of the
solution through the Euclidean method, where the Euclidean time $\tau$ is
imaginary, periodic of period $\beta$ and it is related to the temperature via
$\beta=T^{-1}$. The Euclidean action $I_{{\tiny {{\mbox{Euc}}}}}$ is related
with the free energy $F$ by
\begin{align}
I_{{\tiny {{\mbox{Euc}}}}}=\beta\,F=\beta\left(  M-T\mathcal{S}\right)  ,
\label{freeNRJ}%
\end{align}
where $M$ is the mass and $\mathcal{S}$ is the entropy. On the other hand, in
order to display the boundary term $B$ that will ensure the finiteness of the
Euclidean action, it is enough to consider the following class of Euclidean
metric
\[
ds^{2}=N^{2}(r)f(r)d\tau^{2}+\frac{dr^{2}}{f(r)}+ r^{2} d\Sigma_{\gamma}^{2}
\,.
\]

With the Euclidean time $\tau\in[0,\beta]$, the radial coordinate $r\in
[r_{h},\infty[$ and $Vol(\Sigma_{\gamma})$ standing for the volume of the
three dimensional Euclidean manifold $\Sigma_{\gamma}$, the reduced action
principle reads
\begin{align}
\label{redaction}I_{{\tiny {{\mbox{Euc}}}}}= \beta\,Vol(\Sigma_{\gamma}) \int
N(r)\mathcal{H}(r)\ dr + B\, ,
\end{align}
where the reduced Hamiltonian is given by
\begin{align}
\label{redham}\mathcal{H} = \left(  \frac{5(f-\gamma)^{4}a_{5}}{2r^{6}}-
\frac{2(f-\gamma)^{3} a_{4}}{r^{4}} - \frac{3(f-\gamma)^{2} a_{3}}{r^{2}} -
12(f-\gamma) a_{2} + 3r^{2} \right)  f^{\prime}\nonumber\\
- \frac{3(f-\gamma)^{5} a_{5}}{r^{7}} + \frac{2(f-\gamma)^{4} a_{4}}{r^{5}}
+\frac{2(f-\gamma)^{3} a_{3}}{r^{3}} + 6r(f-\gamma) + 2\Lambda r^{3} \,.
\end{align}

The boundary term is determined requiring that the reduced action
(\ref{redaction}) has an extremum, that is, $\delta I_{{\tiny {{\mbox{Euc}}}}%
}=0$ within the class of fields considered here \cite{Regge:1974zd}. This last
condition implies that
\begin{align*}
\delta B= -\beta\,Vol(\Sigma_{\gamma})\,N\,\left(  \frac{5(f-\gamma)^{4}a_{5}%
}{2r^{6}}- \frac{2(f-\gamma)^{3} a_{4}}{r^{4}} - \frac{3(f-\gamma)^{2} a_{3}%
}{r^{2}} - 12(f-\gamma) a_{2} + 3r^{2} \right)  \delta f \,,
\end{align*}
where the variation is taken between the horizon and infinity.

Varying the reduced action with respect to the fields $N$ and $f$ gives the
following equations:
\begin{subequations}
\begin{align}
\mathcal{H}(r)=0\\
\left(  -\frac{5(f-\gamma)^{4} a_{5}}{2r^{6}} + \frac{2(f-\gamma)^{3})a_{4}%
}{r^{4}} + \frac{3(f-\gamma)^{2} a_{3}}{r^{2}} + 12(f-\gamma)a_{2} -3r^{2}
\right)  N^{\prime}(r) = 0
\end{align}
From the first equation we recover (\ref{wheeler}). In the second equation,
since the couplings $a_{2},a_{3},a_{4},a_{5}$ are generic, one must have
$N^{\prime}(r)=0$, and then, without loss of generality, we can set
$N(r)\equiv1$.

The temperature is fixed requiring regularity of the metric at the horizon
yielding in this case to
\end{subequations}
\[
\beta f^{^{\prime}}(r)|_{r_{h}} = 4\pi,
\]
and using (\ref{wheeler}) we have
\begin{equation}
T = \dfrac{1}{2\pi r_{h}} \left(  \frac{-3a_{5}\gamma^{5} -2a_{4}\gamma
^{4}r_{h}^{2} +2a_{3}\gamma^{3}r_{h}^{4} +6\gamma r_{h}^{8}-2\Lambda
r_{h}^{10}}{5a_{5} \gamma^{4} + 4a_{4}\gamma^{3} r_{h}^{2} -6a_{3} \gamma^{2}
r_{h}^{4} + 24 a_{2} \gamma r_{h}^{6} + 6 r_{h}^{8}} \right)  \,.
\end{equation}

Now we are in position to compute the boundary term $B= B(\infty) - B(r_{h})$.
At the horizon we have $f(r_{h})=0$ and $\delta f|_{r_{h}} = - f^{\prime
}(r_{h})\,\delta r_{h}$. In this situation the variation reads
\begin{align*}
\delta B\big|_{r_{h}} = 4\pi\, Vol(\Sigma_{\gamma}) \left(  \frac{5\gamma^{4}
a_{5}}{2r_{h}^{6}} + \frac{2\gamma^{3} a_{4}}{r_{h}^{4}} - \frac{3\gamma^{2}
a_{3}}{r_{h}^{2}} + 12\gamma a_{2} + 3r_{h}^{2}\right)  \,\delta r_{h},
\end{align*}
which can be trivially integrated as
\begin{align}
B(r_{h}) = Vol(\Sigma_{\gamma}) \left(  - \frac{2\pi\gamma^{4} a_{5}}%
{r_{h}^{5}} - \frac{8\pi\gamma^{3} a_{4}}{3r_{h}^{3}} + \frac{12\pi\gamma^{2}
a_{3}}{r_{h}^{2}} + 48\pi\gamma a_{2} r_{h} + 4\pi r_{h}^{3} \right)  \,.
\end{align}

On the other hand, we can use again (\ref{wheeler}) to see that the variation
at infinity is simply
\begin{align}
\delta B\big|_{\infty} = \frac{\beta\, Vol(\Sigma_{\gamma})}{2} \delta
\mu\,,\nonumber
\end{align}
so that the contribution at the infinity is given by
\begin{align}
B(\infty)  &  = \frac{\beta\, Vol(\Sigma_{\gamma})}{2} \mu\nonumber\\
&  = -\frac{\beta\, Vol(\Sigma_{\gamma})}{2} \left(  -\frac{\gamma^{5} a_{5}%
}{r_{h}^{6}} - \frac{\gamma^{4} a_{4}}{r_{h}^{4}} + \frac{2\gamma a_{3}}%
{r_{h}^{2}} - 12\gamma^{2} a_{2} - 6\gamma r_{h}^{2} + \Lambda r_{h}^{4}
\right)  \,.
\end{align}

Finally, the comparison between the boundary term $B$ and (\ref{freeNRJ}),
allows us to identify the entropy and the mass of the system as
\begin{align}
\mathcal{S}  &  = Vol(\Sigma_{\gamma}) \left(  4\pi
r_{h}^{3}+ 48\pi\gamma a_{2} r_{h} +
\frac{12\pi\gamma^{2} a_{3}}{r_{h}^{2}}- \frac{8\pi\gamma^{3} a_{4}}{3r_{h}^{3}} - \frac{2\pi
\gamma^{4} a_{5}}{r_{h}^{5}}    \right)  \,.\\
\label{mass}M  &  = \frac{Vol(\Sigma_{\gamma})}{2} \left(  \frac{\gamma^{5} a_{5}}%
{r_{h}^{6}} + \frac{\gamma^{4} a_{4}}{r_{h}^{4}} - \frac{2\gamma a_{3}}%
{r_{h}^{2}} + 12\gamma^{2} a_{2} + 6\gamma r_{h}^{2} - \Lambda r_{h}^{4}
\right)
\end{align}

It's now trivial to show that the first law $dM=TdS$ is fulfilled.

As a final comment, we want to stress that the expression for the mass
(\ref{mass}) coincides with the one obtained through the quasilocal
generalization of the ADT formalism \cite{Abbott:1981ff} as presented in Refs.
\cite{Kim:2013zha,Gim:2014nba}.

\section{No-ghosts on AdS}

Linearizing fourth-order gravity theories around maximally symmetric
backgrounds in general leads to ghost degrees of freedom. Some exceptions that
non-trivially avoid this problem are New Massive Gravity as well as
quasi-topological gravities. The former exploits the fact that General
Relativity does not propagate local bulk degrees of freedom in $2+1$%
-dimensions, and the linearized equations around flat space lead to the
massive Fierz-Pauli equation \cite{Bergshoeff:2009hq}. The mechanism that
exorcizes the ghost on AdS in quasi-topological gravities is different. It was
shown in \cite{MyersRobinson} and \cite{originalquartic} that around maximally
symmetric backgrounds, cubic and quartic quasi-topological gravities lead to
the same propagator than G.R. with an effective Newton's constant that
depends on the values of the couplings. Requiring the positivity of this
effective gravity coupling provides an important restriction on the values
that the couplings can take. Now we focus on the linearization of quintic
quasi-topological gravity, given by the action (\ref{fullaction}), around a
maximally symmetric solution with a dressed curvature $\lambda/6$, restricted
by the polynomial $P\left[  \lambda\right]  =0$ in (\ref{PTLambda}).

Recently, the authors of \cite{BuenoCano1} and \cite{BuenoCano2} introduced a
simple method for linearizing higher-derivative theories around maximally
symmetric backgrounds. Their method relies on the evaluation of the Lagrangian
on a deformed curvature that depends on two auxiliary parameters. Considering
derivatives of the effective action with respect to the mentioned parameters,
one can obtain the linearized field equations. The potentially dangerous terms
contain second derivatives of the linearized Einstein tensor as well as of the
linearized Ricci scalar. Such terms will be present for generic combinations
of the curvature invariants. Nevertheless, for quintic quasi-topological
gravity (\ref{fullaction}) one has that the linearized equations around a
constant curvature background of dressed curvature $\lambda/6$ fixed by
(\ref{PTLambda}), read\footnote{Note that for simplicity, we have scaled the
curvature of the background w.r.t. that of \cite{BuenoCano1} as $\lambda
\rightarrow\lambda/6$.}%
\begin{equation}
\frac{dP\left[  \lambda\right]  }{d\lambda}G_{\mu\nu}^{L}=0\ ,
\end{equation}
that implies%
\begin{equation}
\frac{1}{2}\left(  1+\frac{2}{3}a_{2}\lambda-\frac{1}{36}a_{3}\lambda
^{2}+\frac{1}{324}a_{4}\lambda^{3}+\frac{5}{7776}a_{5}\lambda^{4}\right)
G_{\mu\nu}^{L}=0\ ,\label{linealizada}
\end{equation}
where the linearized Einstein tensor is defined by%
\begin{align*}
G_{\mu\nu}^{L}  &  =\bar{\nabla}_{(\mu|}\bar{\nabla}_{\sigma}h_{\ |\nu
)}^{\sigma}-\frac{1}{2}\bar{\square}h_{\mu\nu}-\frac{1}{2}\bar{\nabla}_{\mu
}\bar{\nabla}_{\nu}h+5\Lambda h_{\mu\nu}-\Lambda h\bar{g}_{\mu\nu}\\
&  -\frac{1}{2}\bar{g}_{\mu\nu}\left(  \bar{\nabla}^{\alpha}\bar{\nabla
}^{\beta}h_{\alpha\beta}-\bar{\square}h-4\Lambda h\right)  -4\Lambda h_{\mu
\nu}\ .
\end{align*}

The first term in the parentheses of (\ref{linealizada}) comes from the Einstein-Hilbert term. The
positivity of this effective Newton's constant has to be used to restrict the
values of the couplings.

Note also that if we linearize around a maximally symmetric background that
has a curvature that is a root of $P\left[  \lambda\right]  =0$ with
multiplicity greater or equal to 2, the linearized field equations would
identically vanish. This feature is also well known in Lovelock theories when
vacua coincide (see e.g. \cite{gravitywithoutgraviton}).

\section{Simple solutions for special values of the couplings}

\subsection{Quasi-topological gravity with a unique vacuum}

As in Lovelock theory, one can see that if the couplings are such that all the
roots of the polynomial $P\left[  \lambda\right]  $ in (\ref{PTLambda})
coincide, Wheeler's polynomial can be solved in a simple manner \cite{BHS1},
\cite{BHS2}. For quasi-topological gravity, such degenerate case is obtained
by setting the coefficients as%
\begin{equation}
\Lambda=-\frac{6}{5l^{2}},\ a_{2}=l^{2},\ a_{3}=-6l^{4},\ a_{4}%
=6l^{6},\ a_{5}=\frac{6}{5}l^{8}\ ,
\end{equation}
where $l$ is an arbitrary length scale. In this case the metric can be found
explicitly and yields%

\begin{equation}
ds^{2}=-\left(  \frac{r^{2}}{l^{2}}+\gamma-\mu r^{6/5}\right)  dt^{2}%
+\frac{dr^{2}}{\frac{r^{2}}{l^{2}}+\gamma-\mu r^{6/5}}+r^{2}d\Sigma_{\gamma
}^{2}\ , \label{bhslike}%
\end{equation}
where $\mu$ is an integration constant. This spacetime is asymptotically
locally (A)dS, which can be seen from the asymptotic form of the Riemann
tensor at infinity%
\begin{equation}
R_{\ \ \alpha\beta}^{\mu\nu}=-\frac{1}{l^{2}}\delta_{\alpha\beta}^{\mu\nu
}+p_{\alpha\beta}^{\mu\nu}\mathcal{O}\left(  r^{-4/5}\right)  \ ,
\end{equation}
where $p_{\alpha\beta}^{\mu\nu}$ is a tensor that has constant entries on the
coordinates used for the line element (\ref{bhslike}).

Note also that the solution (\ref{bhslike}) can be obtained as a kind of
dimensional continuation of the black hole solution of Lovelock theory with a
unique vacuum. Indeed in such case the lapse function reads%
\begin{equation}
f\left(  r\right)  =\frac{r^{2}}{l^{2}}+\gamma-\frac{\mu}{r^{\frac{D-2k-1}{k}%
}}\ , \label{bhs}%
\end{equation}
where $D$ is the spacetime dimension and $k$ is the maximum power in the
curvature that appears in the Lagrangian. Lovelock gravity of order $k$
contributes to the field equations provided $k\leq\left[  \frac{D-1}%
{2}\right]  $, where $\left[  \cdot\right]  $ stands for the integer part.
Nevertheless, continuing the solution (\ref{bhs}) to $D=5$ and $k=5$,
correctly reproduces the line element (\ref{bhslike}). For cubic
quasi-topological gravity, this property was already pointed out in \cite{OR}.

\subsection{Pure quasi-topological gravity}

Another simple explicit solution arises when considering the quintic term plus
a cosmological term in the action. In analogy to what happens with Lovelock
gravity, we call this case the pure quasi-topological case. In the context of
Lovelock theory, this case has received attention since it's the simplest case
that admits a unique constant curvature solution that also propagates a
graviton \cite{Cai:2006pq}-\cite{Concha:2016kdz}. The
theory therefore reads%
\begin{equation}
I_{PQG}=\int\sqrt{-g}d^{5}x\left[  -2\Lambda+a_{5}\mathcal{L}_{5}\right]  \ ,
\end{equation}
and the black holes solution reduces to%
\begin{equation}
ds^{2}=-\left(  \gamma-\left(  \frac{\left(  \Lambda r^{4}+\mu\right)  r^{6}%
}{a_{5}}\right)  ^{1/5}\right)  dt^{2}+\frac{dr^{2}}{\gamma-\left(
\frac{\left(  \Lambda r^{4}+\mu\right)  r^{6}}{a_{5}}\right)  ^{1/5}}%
+r^{2}d\Sigma_{\gamma}^{2}\ .
\end{equation}
It's interesting enough to note that even in the absence of an
Einstein-Hilbert term, the asymptotic behavior of the lapse function emulates
that of GR, since at infinity%
\begin{equation}
f\left(  r\right)  =-\left(  \frac{\Lambda}{a_{5}}\right)  ^{1/5}r^{2}%
+\gamma-\frac{\tilde{\mu}}{r^{2}}+\mathcal{O}\left(  r^{-6}\right)  \ .
\end{equation}
The limit $\Lambda\rightarrow0$ can be taken in this solution which provides
an asymptotically locally flat black hole only for the hyperbolic case
$\gamma=-1$.

\bigskip

\section{Conclusions and final remarks}

We have presented a new quasi-topological theory in five dimensions, that is
quintic in the Riemann tensor. The theory is ghost-free around AdS and the
linearized equations around a constant curvature background reduce to the
linearized Einstein equations with an effective Newton's constant. The theory
is defined by five dimensionful coupling constants, and the positivity of the effective
Newton's constant of the fluctuations on AdS has to be imposed as a
restriction on the space of couplings. The theory also admits black hole
solutions which can be integrated exactly up to the solution of a quintic
algebraic equation, that due to Abel-Rufini theorem, cannot be solved in terms
of radicals. Requiring the existence of an event horizon, we have been able to
compute the temperature, mass and entropy of the black hole solutions. These
black holes provide for finite temperature duals of a CFT living at the
boundary of AdS spacetime.

It would be interesting to construct $\mathcal{R}^{5}$ quasi-topological
gravities in arbitrary dimensions. From what we have learned from the cubic
\cite{OR} and quartic cases \cite{originalquartic}, it is natural to
conjecture that this theory might exist for all dimensions lower than ten, and
from dimension 11 it might reduce to Lovelock theory plus a combination of
invariants whose contribution cancels when evaluated on spherically symmetric spacetimes.

Cubic and quartic quasi-topological gravities have been explored in many
different directions. Holographic and unitarity studies including $\eta/s$ and
central charges of the dual CFTs were done in \cite{holoqtg3}%
-\cite{unitarityqtg}. The properties of holographic superconductors were
explored in \cite{holosuperconducSiani}-\cite{holosuperconduc1}. Exact
solution containing matter fields were constructed in \cite{Lifshitzqtg}%
-\cite{magneticbraneqtg} including also scalars conformally coupled to the
quasi-topological densities \cite{scalarhair}. The thermodynamics in the
extended phase space for these theories was studied in
\cite{extendedphasespaceqtg1} and \cite{extendedphasespaceqtg2}. It would be
interesting to see if the results found in those references is affected by the
presence of the quintic quasi-topological term, or whether the results are
generic for this family of theories.

As mentioned above, besides the clear structure provided by the family of
theories introduced in \cite{OR}, of order $k$ in dimension $D=2k-1$, the
general form of quasi-topological gravities of arbitrary high order in five
dimensions, is not clear. For the quartic case for example, one might be
tempted to consider combinations of the invariants:
\begin{align*}
B_{1}  &  =\delta_{b_{1}...b_{8}}^{a_{1}...a_{8}}C_{\ \ a_{1}a_{2}}%
^{b_{1}b_{2}}...C_{\ \ a_{7}a_{8}}^{b_{7}b_{8}},\ B_{2}=\delta_{b_{1}...b_{8}%
}^{a_{1}...a_{8}}R_{\ \ a_{1}a_{2}}^{b_{1}b_{2}}C_{\ \ a_{3}a_{4}}^{b_{3}%
b_{4}}...C_{\ \ a_{7}a_{8}}^{b_{7}b_{8}},\\
B_{3}  &  =\delta_{b_{1}...b_{8}}^{a_{1}...a_{8}}R_{\ \ a_{1}a_{2}}%
^{b_{1}b_{2}}R_{\ \ a_{3}a_{4}}^{b_{3}b_{4}}C_{\ \ a_{5}a_{6}}^{b_{5}b_{6}%
}C_{\ \ a_{7}a_{8}}^{b_{7}b_{8}},\ B_{4}=\delta_{b_{1}...b_{8}}^{a_{1}%
...a_{8}}R_{\ \ a_{1}a_{2}}^{b_{1}b_{2}}R_{\ \ a_{3}a_{4}}^{b_{3}b_{4}%
}R_{\ \ a_{5}a_{6}}^{b_{5}b_{6}}C_{\ \ a_{7}a_{8}}^{b_{7}b_{8}}\\
B_{5}  &  =\delta_{b_{1}...b_{8}}^{a_{1}...a_{8}}R_{\ \ a_{1}a_{2}}%
^{b_{1}b_{2}}R_{\ \ a_{3}a_{4}}^{b_{3}b_{4}}R_{\ \ a_{5}a_{6}}^{b_{5}b_{6}%
}R_{\ \ a_{7}a_{8}}^{b_{7}b_{8}}\ .
\end{align*}
Substituting the Weyl tensor in terms of the Riemann tensor and its traces,
one can see that the combination%
\begin{equation}
\mathcal{B}=\frac{1}{\left(  D-5\right)  }\sum_{i=1}^{5}\xi_{i}B_{i}\ ,
\end{equation}
is well behaved for a general metric in $D=5$ if and only if%
\begin{align*}
\xi_{3}  &  =-6\xi_{1}-3\xi_{2}\ ,\\
\xi_{4}  &  =26\xi_{1}+15\xi_{2}\ ,\\
\xi_{5}  &  =-21\xi_{1}-13\xi_{2}\ .
\end{align*}
Then, in order to remove the higher derivative contributions to the field
equations one would have to engineer additional terms that on spherical
symmetry contribute as those in the combinations $\mathcal{B}$. Besides the
independent terms $Tr_{\left(  p\right)  }\left(  C^{4}\right)  $, one might
also consider $R Tr_{\left(  1\right)  }\left(  C^{3}\right)  $ and
$R Tr_{\left(  2\right)  }\left(  C^{3}\right)  $.

\section{Acknowledgement}

We thank Gaston Giribet and Sourya Ray for enlightening discussions and
collaboration on similar topics. This work is partially supported by FONDECYT
grants 1130423, 1141073 and 3150157. This project is also partially funded by
Proyectos CONICYT- Research Council UK - RCUK-DPI20140053. A.C. and J. O.
would like to thank the International Center for Theoretical Physics (ICTP),
Trieste, Italy, where part of this work was carried out. J.O. thanks also the
support of ICTP Associates program.

\section{Appendix}

The coefficients that define the new quintic quasi-topological in equation (\ref{uff}) are:%
\begin{align*}
A_{1}  &  =\frac{9497}{17767320},\ A_{2}=-\frac{759299}{71069280}%
,\ A_{3}=\frac{124967}{5922440},\ A_{4}=\frac{759299}{23689760},\\
A_{5}  &  =\frac{197761}{11844880},\ A_{6}=\frac{1362599}{23689760}%
,\ A_{7}=-\frac{5006573}{11844880},\ A_{8}=-\frac{9290347}{71069280},\\
A_{9}  &  =\frac{3400579}{11844880},\ A_{10}=-\frac{6726521}{11844880}%
,\ A_{11}=\frac{363777}{23689760},\ A_{12}=\frac{6348187}{47379520},\\
A_{13}  &  =-\frac{9487667}{71069280},\ A_{14}=-\frac{6454201}{8883660}%
,\ A_{15}=-\frac{34697591}{142138560},\ A_{16}=-\frac{5643853}{71069280},\\
A_{17}  &  =-\frac{29094011}{71069280},\ A_{18}=-\frac{48458099}%
{71069280},\ A_{19}=\frac{1547591}{740305},\ A_{20}=\frac{78763919}%
{71069280},\\
A_{21}  &  =-\frac{10718341}{17767320},\ A_{22}=\frac{9629717}{17767320}%
,\ A_{23}=\frac{1113473}{17767320},\ \ A_{24}=-\frac{16111757}{17767320}.
\end{align*}

\end{document}